\documentstyle[twoside]{memsait}

\begin{opening}
\title{CBS - A PROGRAM FOR CLOSE BINARY SYSTEM LIGHT CURVE ANALYSIS }
\author{L. SOLMI$^1$, M. GALLI$^2$}
\institute{$^1$Universit\`a di Bologna, Dipartimento di Astronomia, Via
Zamboni 33, Bologna, Italy \\$^2$ENEA, Area Energia e Innovazione,
Dipartimento Sviluppo Tecnologie di Punta,\\ C.R.E. Bologna, Viale Ercolani 8,
40138 Bologna, Italy}
\date{}
\end{opening}

\begin{document}

\oddpagefooter{\sf Mem. S.A.It., Vol. XX, 1993}{}{\thepage}
\evenpagefooter{\thepage}{}{\sf Mem. S.A.It., Vol. XX, 1993}
\

\bigskip

\begin{abstract}



CBS is a new program for binary system light curve analysis,
it generates synthetic light curves for a binary system, accounting for
eclipses, tidal distortion, limb darkening, gravity darkening and  reflection;
it is also possible to compute the light contribution and eclipses
of an accretion disk. The bolometric light curve is generated,
as well as curves  for the U,B,V,R,I colour bands.
In the following we give a brief description of the first version of the
program and show some preliminary results.
\end{abstract}
\section{Introduction}
For many years the photometric analysis of  binary sys\-tems has been
performed by the rectification method developed by Russel (1912, 1952);
with this method the light curve is analysed in terms of the eclipses
of two limb darkened spherical disks by removing
off-eclipse effects such as reflection and tidal distortion.
The Russel method was
improved by different authors, using an analytical approach (Kopal 1959),
but the detailed computations of close binary systems
light curves, with strong
tidal and reflection effects, can
be done only with computers, by  numerical calculation.
\par
\smallskip
In the seventies, when powerful com\-puters became
available to astronomers, many light synthesis programs where written
by different authors, such as
Wilson and Devinney (1971),
Wood (1971), Lucy (1968), Linnel (1984) and others.
The most popular of these is the Wilson-Devinney pro\-gram,
which represents the stars, deformed by ti\-dal and rotational
forces, by means of equipotential Roche surfaces,
 assuming central
condensation, synchronous rotation and circular orbits.
The gravity darkening effect is considered.
Reflection is treated in an approximate way:
the geometry is simplified by con\-sider\-ing the irradiating surface
as a point source, and a term is introduced in the
albedo coefficient to account for its shape.
For the eclipse computation the surfaces of the two stars are represented by a
number of surface elements ( usually about thousand ); for each phase the
star closer to the observer is analysed first
and an analytical approximation for  the boundary of the visible
part of the star is computed from the boundary surface elements.
This is used to exclude the eclipsed part of the second star
when summing the light of all the visible surface elements.
The differential correction method is used to obtain the
best parameters for the light curve.
\par
\smallskip
The Wilson Devinney program gives a satisfactory solution for the light curves
of most close binary systems; but most of the modern interest in close binary
systems is focused on accretion disks fenomenology and stars with a collapsed
companion. A Wilson type approach can't handle accretion disks and
complex geometrical configurations;
 in fact only few attempts have been done up to now in representing the
disk contribution to the light curve (Wilson, Caldwell 1978; Horne 1985).
\par
\smallskip
In order to consider the contribution of hot spots and accretion disks
we have written a new light synthesis program, which uses a different
approach, suited to handle complex geometrical configurations.
We are still developing the program; the present version isn't
optimized, doesn't treat overcontact bi\-naries and hot spots,
doesn't contain a minimum finding  algorithm to look for the best parameters;
all these features will be included in a future version.
%
\section{Program description}
%
The CBS program computes the light curve for a system
consisting of two stars and an accretion disk; the two stars can be
represented by spheres or roche equipotential surfaces, the disk
as a thick ring.
\par
\smallskip
The stars and the disk are described by a number of surface elements;
for each of them the tem\-pera\-ture is ob\-tained by a gravity darkening law,
the emitted bolometric luminosity is computed by assuming a Plank law.
Also the  position, orien\-ta\-tion and approximate extension of each
surface element are stored in memory.
\par
\smallskip
The reflection effect is treated following the vector meth\-od of
Chen and Rein (1969), considering multiple reflections and the full
geometry of the system. The computing time
increases here as the square of the number of surface elements,
but a very detailed calculation isn't necessary for reflection;
for this reason we group more surface elements together into a "coarse
surface element", used only in this part of the program.
To calculate the reflected light
we consider all the possible light paths between two
coarse surface elements of the two stars which don't intercept
the disk and all the paths between a star and the disk which don't
intercept the other star; a transmission function is computed for each path,
including  linear limb darkening  and an input-given albedo coefficient.
These functions are used to compute the energy transmitted along each path
and a reflected luminosity for each surface element, which
is added to the bolometric one.
\par
\smallskip
The light received by the observer at each phase is obtained by
projecting the surface elements into the visual plane.
The visual plane is divided into a number of cells;
a surface element is eclipsed if it falls into the same cell of another surface
element which is closer to the observer.
The case of more than one surface element of the same object
falling into the same cell and the case of a surface element with a projected
dimension greater than a single cell are properly treated.
The received light is obtained by summing over all the non-eclipsed
surface elements, assuming a linear limb darkening law.
\par
\smallskip
This approach to the eclipse computation is not limited to treat a
particular geometry and can be used for disks with hot spots,
spheres and Roche lobe stars without any change.
Its disadvantage is the high number of surface elements needed to
represent the objects. To obtain more than about 1\% accuracy in the
eclipse computation we need about 10000 surface elements for each object,
this means a computer memory of some tenths of Mbytes, a common value for many
modern workstations.
The present version of the program computes the light curve of an Algol type
binary, with 100 phase values,  in about 10 minutes on a Microvax 3300
computer using 15000 surface elements for each star.
A forthcoming  optimized version will give better performances.
\par
\smallskip
The input of the program consists of the pole temperature of the stars,
the limb, gravity darkening and albedo coefficients, the orbit inclination
and some parameters describing the system geometry as the presence of the disk,
 its height and radius,  the Roche potentials etc.
A number of run parameters of the program, as the number of surface elements,
the number of visual plane cells, the normalization par\-ameters etc.
can also be given by the user.
\par\smallskip
In fig.1 we show the light curve of IM-Aur (BV 267), an example of
a semi-detached Algol type variable.
 The system parameters are those
obtained by Rafert (1990);
for the first star:
$ T=12000 \;\; \Omega=3.103  x=0.75 \;\; g=0.63 \;\; A=1 \;\; ;$
for the second:$ T=5945 \;\; \Omega=2.4913  x=0.675 \;\; g=0.32 \;\;
A=0.8 \;\;. $   $ i=75.20^o \;\; q=0.3114 . $
We didn't put disk or hot spots.
The continuous line is the monochromatic
light curve at $5410 A $, obtained by CBS;
the dashed line is obtained by the Wilson-Devinney program and
the dots are observations from Margoni et al. (1966).
The difference between the two synthetic light curves is due to the
different treatment of reflection. The overall agreement with
observations is good.
\par\smallskip
In fig.2 the effects of reflection and limb darkening are shown: the dashed
line is a light curve obtained without reflection contribution,
the dotted line without limb darkening.
\acknowledgements
We are grateful to Professor C. Bartolini, for comments, discussion and moral
support, and to Doctor M. Lolli for discussions and the data
files to test the program.
%

%
\begin {figure} [tbph]
\vspace {8cm}
\includegraphics{figura1.ps}
\caption [1] {IM-Aur: the CBS light curve (continous), the Wilson light curve
(dashed) and the data points (dots).}
\end {figure}
\begin {figure} [tbph]
\vspace {8cm}
\includegraphics{figura2.ps}
\caption [2] {IM-Aur: the CBS light curve (continous), without reflection
(dashed) and without limb darkening (dotted).}
\end {figure}
\end{document}